\documentstyle{llncs}
\begin{document}

\title{Quantum Key Distribution and String Oblivious Transfer 
       in Noisy Channels}

\author{Dominic Mayers\thanks{Supported in part by NSERC \& FCAR}}

\institute{DIRO,Universit\'e de Montr\'eal, C.P.~6128,\\
           Succursale CENTRE-VILLE, Montr\'eal (Qu\'ebec), Canada H3C 3J7\\
           and\\
           Computer Science Department,\\
           Maharishi University of Management,\\
           Fairfield, Iowa 52557-1121, U.S.A.}

\maketitle

\begin{abstract}
We prove the unconditional security of a quantum key
distribution (QKD) protocol on a noisy channel 
against the most general attack allowed by quantum physics. 
We use the fact that in a previous paper we have reduced
the proof of the unconditionally  security of this QKD 
protocol to a proof that a corresponding Quantum String Oblivious 
Transfer (String-QOT) protocol would be unconditionally 
secure against Bob if implemented on top of an unconditionally 
secure bit commitment scheme.  We prove a lemma that 
extends a security proof given by Yao for a (one bit) 
QOT protocol to this String-QOT protocol. This result
and the reduction mentioned above implies the unconditional 
security of our QKD protocol despite our previous proof
that unconditionally secure bit commitment schemes are impossible.
\end{abstract}

\section{Introduction and Brief History}
One of the most popular 
application of quantum physics to cryptography is 
quantum key distribution (QKD).
In an ideal QKD, Alice and Bob who share no secret information
initially, share a secret string $s$ at the end.
An eavesdropper, typically called Eve, should learn 
nothing about the secret string $s$, except perhaps
for its length.  
  
In this paper, we prove the security of a QKD protocol
against the most general attack  allowed by quantum physics.
This QKD protocol works with a noisy quantum channel, an imperfect
measuring apparatus, but requires 
a perfect source
and 
a faithful classical channel.
A channel is {\em faithful} if no one can modified
a message sent in the channel without being detected.  
The need for a faithful classical channel is not a problem because
a secret string $s_0$ initially shared between Alice and
Bob can be used to simulate a faithful classical channel
by use of an unconditionally secure classical authentication scheme
\cite{wc81}.  
We assume a perfect source to avoid the technical
difficulty associated with many photons per pulse.  

Our preliminary version of the protocol 
uses a random linear code for error correction.
Random linear codes are very difficult
to decode.  However, this problem can be solved and
a version of the protocol using an 
efficient error correcting code and with no requirement 
for a perfect source will be considered in the journal 
version of this paper. 

In addition to QKD, other applications of quantum physics to 
cryptography have been proposed.  The most popular are quantum bit
commitment (QBC) and quantum oblivious transfer (QOT).  
We briefly review these  protocols since we
shall refer to them in our results. 
In the bit commitment task from Alice to Bob, Alice commits a bit $b$.
Later, if Bob asks Alice to {\em unveil} the
commitment,  he receives the bit $b$. 
The main point is that Alice cannot change the value of $b$ 
and Bob learns nothing about $b$ unless Alice unveils it.    
In the oblivious transfer task from Alice to Bob, Alice enters a 
bit $b$,  Bob receives a perfectly random bit
$c$ and he learns the value of $b$ if and only if $c = 0$.  Alice learns
nothing about~$c$.

The first quantum bit commitment protocol ever proposed is due to 
Bennett and Brassard \cite{bb84}.  The authors themselves knew at the time that
this protocol is insecure.   Other quantum bit commitment protocol 
have been proposed, but none of them could be
proven unconditionally secure.  In fact, it has been shown  recently
that unconditional security for quantum bit commitment is 
impossible~\cite{mayers95b,mayers96a,mayers96b}.  A proof of
computational security for a quantum bit commitment 
protocol is still possible, 
but none is currently available.  The absence of a provably
secure  bit commitment is  unfortunate because
all the known quantum oblivious transfers  are built on
top of bit commitment, that is, they use quantum bit commitment
as a sub-protocol.    

The first quantum oblivious transfer protocol which would be secure 
if implemented on top of a secure bit commitment protocol
has been proposed by Cr\'epeau~\cite{crepeau90}.
Its security against most but not all  reasonable attacks 
allowed by the current technology  has been shown 
in \cite{bbcs92}.  The first proof that considered the most general 
attack allowed by quantum physics, including the so called
{\em coherent} measurements on many photons at a time,  has 
been obtained by Yao~\cite{yao95}.   
Yao's proof is an important step and provides useful techniques, 
but it provides no security because, as for all the
previous proofs \cite{bbcs92,ms94},  
it requires a secure bit commitment
and none has yet been proven secure.

Now, we are back to  QKD. The security of a QKD protocol against most 
but not all reasonable attacks allowed by the current technology 
has been established in \cite{bb89,bbbss90}.  
In~\cite{mayers95a}, we have reduced the unconditional security of any 
QKD protocol of a certain kind to a proof that a corresponding 
String-QOT protocol would be unconditionally secure if implemented
on top of an unconditionally secure bit commitment scheme.
A QKD protocol of the appropriate type is associated
with a corresponding String-QOT protocol.
The standard QOT protocol in Yao's proof
turns out to be associated with a QKD protocol of
the appropriate type. Therefore, the unconditional security of this 
QKD protocol is obtained from the above reduction.
However, there are two problems with this protocol.
First, the QOT protocol in Yao's proof is a standard one bit QOT, 
therefore only one secret bit is returned in the QKD version.
One can repeat the protocol $n$ times to obtain a secret string of length $n$,
but an initial secret key $s_0$ is required to simulate a 
faithful classical channel and, therefore, each execution of the 
protocol uses more secret bits than it returns back!
Second, the QOT protocol in Yao's proof, and thus the corresponding
QKD protocol, requires a noiseless quantum channel and 
a perfect source.

In this paper, to pursue the original idea of~\cite{mayers95a}, 
we extend Yao's proof to a String-QOT protocol
associated via the above reduction with a ``strong''  QKD protocol.
Therefore, we have the unconditional security of this QKD
protocol.  This QKD protocol returns a 
secret string $s$ that is longer than the required initial 
string $s_0$.  Also, it works in a noisy quantum channel.  
Note that our proof for this QKD protocol
considers any kind of errors in Bob's apparatus  
because we give full control over both the 
channel and the apparatus to a dishonest Bob in String-QOT.

It is shown in \cite{bcm96} that the security of any OT protocol
implies the security of a String-OT protocol.  In particular,
the security of the QOT protocol in Yao's proof 
implies the security of a String-QOT protocol.
However, the security of the resulting String-QOT protocol 
does not imply the  security of a QKD protocol via the above
reduction because it is not of the required type.  
Yao did not mention the possibility of generalizing
his proof to the String-QOT case.  It should be said that Yao 
was not aware of the above reduction 
(or did not believe it) at the time he wrote his  
paper \cite{yao95}.  Yao has announced in \cite{yao95}
that in the journal version of his paper 
the QOT protocol will work on a noisy channel
but our String-QOT protocol has been designed to work 
on a noisy channel without much additional effort.  

\section{Related results}
The main problem that one must address in the design
of a QKD protocol is that Alice and Bob must exchange
quantum systems, let say photons, and there is no way 
to distinguish interaction of these photons
with the environment and interaction 
of these photons with Eve's measuring
apparatus.  Therefore, Eve can always succeed 
to {\em entangle} her measuring apparatus 
with the exchanged photons without being detected.
Later, if these photons are used to define the
shared key, Eve can
obtain information about this key.
However, using privacy amplification
techniques, one can make this information arbitrarily
small.  
For example, in the QKD protocol considered in this paper,
a classical string $w' \in \{0,1\}^N$ is 
stored in $N$ photons traveling from Alice to Bob.  
Because Eve can obtain
information about $w'$, privacy amplification
must be used to distill from $w'$ a shorter but 
secret string $b = h(w')$.
Privacy amplification is an essential part of
any QKD protocol.  Privacy amplification
in the QOT protocol of Yao's proof corresponds to
the fact that the secret bit is the exclusive or
of all the bits of $w'$.  

Much after the BB84 protocol of~\cite{bb84} have been proposed,
Ekert suggested a scheme in which EPR pairs are created
and the photons in each pair are split between Alice
and Bob \cite{ekert91}.  In this EPR scheme, 
no information is stored in the photons before they are sent,
therefore one would hope that no information can be
extracted by Eve.  However, Eve can still entangle her
apparatus with the photons and  it has been
shown that the kind of attacks that could work against the BB84 
scheme correspond to attacks that would work 
against this EPR scheme~\cite{bbm93}. 
This result highly suggested that EPR pairs might not be useful
for quantum cryptography.

However, recently Deutsch, Ekert and al. proposed another EPR-based protocol
with a new element, an {\em entanglement purification} procedure 
also called in this context a quantum privacy amplification 
procedure~\cite{dejmps96}.  Entanglement purification \cite{bbpssw96} 
allows Alice and Bob to generate, from any supply of pairs of photons
with non-zero entanglement,  a smaller set of maximally
entangled EPR pairs whose entanglement with any outside
system, including Eve's apparatus, is arbitrarily low.    
Deutsch, Ekert and al. reasonably argue that their protocol is
unconditionally secure against the most general attack
allowed by quantum physics.  An interesting point is
that privacy amplification is done at the quantum level,
and one can hope that this kind of privacy
amplification procedure is more efficient. 
On the other hand, working prototypes for protocol
that use simple quantum coding schemes  
already exist~\cite{trt93a,trt93b,mbg93,rot94,hughes96}, 
whereas the technology required for this
EPR-based protocol is not yet available.  

Let us emphasis that in a security proof for a QKD 
or a String-QOT protocol one must consider carefully 
the criteria to reject or accept an execution of the protocol.
This criteria always exists for 
a given lower bound on the length of the shared key
or string.  In the case of our String-QOT protocol, Alice
must detect less than $\delta n$ errors.
One must show that this criteria 
implies that the cheater cannot succeed.
This analysis is difficult in the case of
the most general attack allowed by quantum physics
and to our knowledge only Yao's paper~\cite{yao95} deals 
rigorously with this issue.   

The purpose of quantum cryptography 
is not only to prove the security of protocols. 
We also want to design more efficient protocols and see how 
efficient are these protocols in theory and in practice.   
Biham and Mor have obtained the maximal theoretical efficiency of
the QKD protocol of \cite{b92} against a restricted but still
reasonable type of attacks~\cite{bm96}.  Furthermore,
it is reasonable to believe that we could eventually
prove that the security parameter
required against this restricted
type of attack is not too far from the security parameter
required against the most general attack.

\section{Some algebra}
Typically, a quantum protocol involves many systems and
each system is associated with its own Hilbert space ${\cal H}$ also called
a state space.  For example, the polarization of a photon
is associated with a two dimensional Hilbert space.  
The inner product of ${\cal H}$ evaluated on
$(|\phi\rangle,|\psi\rangle) \in {\cal H}^2$ 
is denoted $\langle \phi | \psi \rangle$.
For every vector $|\phi\rangle \in {\cal H}$,
let $|\phi\rangle^{\dagger}  : {\cal H} \rightarrow \bbbc$ be
be the linear functional on ${\cal H}$
which, when evaluated on any vector
$|\psi\rangle \in {\cal H}$, simply returns
the inner product $\langle \phi | \psi\rangle$.
For obvious reason, $|\phi \rangle^{\dagger}$ is more
conveniently denoted $\langle \phi |$.  
In terms of matrices, one represents a vector $|\psi\rangle \in {\cal H}$
as a column matrix.  The operation ``$\dagger$'' on a matrix
is simply the transpose conjugate, therefore $\langle \psi | $ is
represented by a row matrix.

The space of linear functionals on ${\cal H}$ is
denoted ${\cal H}^{\dagger}$.  
It is called the dual of ${\cal H}$.  
The inner product of ${\cal H}$ is also an operation
on the cartesian product  ${\cal H}^{\dagger} \times {\cal H}$.  
This operation can be generalized to any
cartesian product of the form 
${\cal G}_1 \times \ldots \times {\cal G}_n$
where each space ${\cal G}_i$ occurs only once 
and is either a state space ${\cal H}$ or its dual.  
We simply let any functional
$\langle \phi | \in {\cal G}_i = {\cal H}^{\dagger}$ 
operate on the state $| \psi \rangle \in {\cal G}_j = {\cal H}$ 
to its right, if one exists.  Every thing
else should not be simplified.   For example,
consider $|\phi_1\rangle \in {\cal H}_1$,
$\langle \psi_1 | \in {\cal H}_1^{\dagger}$,
$|\phi_2 \rangle \in {\cal H}_2$ and
$\langle \psi_2 | \in {\cal H}_2^{\dagger}$.
We have $\langle \psi_1 | \psi_2 \rangle | \phi_1 \rangle \langle \psi_2 |
= \lambda_1 |\psi_2 \rangle \langle \phi_2 |$ where
$\lambda_1 = \langle \psi_1 | \phi_1 \rangle \in \bbbc$.  
The object $M = |\psi_2\rangle \langle \phi_2|$ cannot be
simplified, but it can operate on other objects.
For instance $M$ on $|\eta_2\rangle \langle \phi_3 | \in {\cal H}_2
\times {\cal H}_3^{\dagger}$ returns 
$ |\psi_2 \rangle \langle \phi_2| \eta_2\rangle
\langle \phi_3| = \lambda_2 |\psi_2 \rangle \langle \phi_3 |$ where
$\lambda_2 = \langle \phi_2 | \eta_2 \rangle \in \bbbc$.

The tensor product ${\cal G}_1
\otimes \ldots \otimes {\cal G}_n$ 
can be interpreted as the span of the 
product  ${\cal G}_1 \times \ldots \times {\cal G}_n$.
If $|\phi_1\rangle |\phi_2\rangle$
and $|\psi_1\rangle |\psi_2 \rangle$ belong to
${\cal H}_1 \times {\cal H}_2$ then the
sum $|\phi_1\rangle |\phi_2\rangle +
|\psi_1\rangle |\psi_2 \rangle$ belongs 
to ${\cal H}_1 \otimes {\cal H}_2$.
A formal definition of this tensor product is usually not 
so enlightening, so none is given
here, but the basic idea is simply to extend by linearity
the operations that are defined above.  Two objects that
cannot be distinguished via these operations (neither as
operators or as operands) are  considered to be identical.
One should notice the following rules:
\begin{itemize}
\item For every ${\cal H}$,
every pair of objects in ${\cal H} \cup {\cal H}^{\dagger}$ does 
not commute, but everything else commute.
\item Because $\langle \phi | \psi \rangle
= \langle \psi | \phi \rangle^*$, where ``$*$''denotes
the complex conjugate,  we have
$|\langle \psi | \phi \rangle |^2 = 
\langle \psi  | \phi \rangle \langle \phi | \psi \rangle
= \langle \phi | \psi \rangle \langle \psi | \phi \rangle$. 
\item For any objects $M_1, \ldots , M_n$, we have
      $(M_1 \ldots M_n)^{\dagger} = M_n^{\dagger} \ldots M_1^{\dagger}$.
      In particular, $(|\psi\rangle\langle \phi|)^{\dagger}
      = |\phi\rangle \langle \psi |$.  
\end{itemize}

The trace of an operator $M \in {\cal H} \otimes {\cal H}^{\dagger}$,
i.e., from ${\cal H}$ into ${\cal H}$, is defined
by ${\rm Tr}(M) = \sum_{\alpha} \langle \psi_{\alpha} | M |
\psi_{\alpha} \rangle$ where $\{ |\psi\rangle_{\alpha} \}$
is any orthonormal basis of ${\cal H}$.  This definition 
is independent of the basis $\{ |\psi_{\alpha}\rangle \}$.  

For $z,z' \in \{0,1\}^n$, $(z \oplus z') \in \{0,1\}^n$ is given
by $(z \oplus z')_i = z_i \oplus z'_i = z_i + z'_i\; ({\rm mod}\; 2)$, 
and $z \odot z'= \oplus_i (z_i \times z'_i)$.  
The set $\{0,1\}$ with the operation $\oplus$ and the ordinary
product is a finite field denoted ${\rm GF}(2)$.  
The set ${\rm GF}(2)^n$ with the operation $\oplus$ is a vector
space over the field ${\rm GF}(2)$.  
Let $f$ be a $m \times n$ boolean matrix and $z$ a boolean 
string of length $n$,  the product $f z$ is the ordinary
matrix operation with the sum modulo $2$ where
$z$ is seen as a boolean column matrix.   

\section{Quantum preliminaries}
The state of a system, 
also called a pure state, is represented by a vector 
$|\psi\rangle$ of norm $1$ in the associated Hilbert space 
${\cal H}$.  The state space of a system made of $n$
subsystems with state spaces ${\cal H}_1, \ldots, {\cal H}_n$
is the tensor product ${\cal H}_1 \otimes \ldots \otimes
{\cal H}_n$.  

A completely refined measurement on ${\cal H}$ 
is a set of outcomes $v$ where every outcome $v$ 
is associated with a vector $|\phi_v\rangle \in {\cal H}$,  
but here the norm could be anything between $0$ and $1$.  
The probability of $v$ given 
the initial state $|\psi\rangle \in {\cal H}$ is simply 
$|\langle \phi_v |\psi\rangle|^2 = \langle \phi_v | 
           \psi\rangle \langle \psi | \phi_v \rangle$.
The only requirement on the states $|\phi_v \rangle$ is
that $\sum_v |\phi_v \rangle \langle \phi_v | = {\bf I}$,
the identity operator.  This is equivalent to say that, for every
initial state $|\psi\rangle$, the sum of the probabilities
over the outcomes $v$ is $1$.  

The final quantum state left after the measurement is
some state $|v\rangle$ which should not be confused with
the vector $|\phi_v\rangle$. 
The operation associated with $v$ is given by 
$M_v = |v\rangle\langle \phi_v |$.  One may check that
the probability of $v$ given the initial state $|\psi\rangle$ 
is $\| M_v |\psi\rangle \|^2$, the square of the norm of $M_v |\psi\rangle$. 
The final state $|v\rangle$ can
be anything because just at the end of the measurement
one is free to store the residual quantum information into the 
final state $|v\rangle$ of his choice.
If $\Omega = \{|\phi_v\rangle\}$ is a basis of ${\cal H}$,
a measurement in the basis $\Omega$ is simply the measurement
that associate $v$ to $|\phi_v\rangle$.  Such a measurement
is called an orthogonal measurement.    

Now, let us generalize to incomplete measurement
the above definition. 
The most general measurement on ${\cal H}$ is a set of outcome $k$
where every outcome $k$ is associated with an 
operator $M_k$ on ${\cal H}$.  The difference
with a complete measurement is that $M_k$
is in general a sum $M_k = \sum_v |v\rangle\langle \phi_v|$
rather than only a rank one operator $M_k = |k\rangle\langle \phi_k|$.
The only requirement on the operators $M_k$
is that $\sum_k M_k^{\dagger} M_k = {\bf I}$. 
The image of $M_k$ 
can be any sufficiently large state space ${\cal H}_k$,
because just at the end of the measurement
one is free to store the residual quantum information 
into the system of his choice. For example, the quantum 
information can be send from the
state space of a photon into the state space of an atom. 
The probability of $k$ given an initial state 
$|\psi\rangle$ is $\| M_k |\psi\rangle \|^2$.

Every measurement ${\bf M}$ on a state space ${\cal H}$
which returns an outcome $k$
can be refined by executing another
measurement ${\bf M}'$ on ${\cal H}_k$. 
The new measurement ${\bf M}'$ may depend upon $k$.
Let $M'_v$ be the operation on ${\cal H}_k$ 
associated with the outcome $v$ of ${\bf M}'$.  
The operation on the original space
${\cal H}$ associated with the overall outcome $(v,k)$ 
is simply $M_{(v,k)} = M'_v M_k$.  

If a quantum preparation contains a pure state $|\psi_{\alpha}\rangle$ with
probability $p_{\alpha}$,  then one may conveniently represent
this preparation by the operator $\rho = \sum_{\alpha} p_{\alpha}\; 
|\psi_{\alpha}\rangle \langle \psi_{\alpha}|$.  
The idea is that the probability of $v$ given the preparation
represented by $\rho$ is simply $\langle \phi_v | \rho | \phi_v \rangle$.  
This works even if the  initial states $|\psi_{\alpha}\rangle$ are not
orthogonal.  Note the important fact that two distinct preparations 
may correspond to a same density operator.  
Even for an incomplete measurement on a given preparation, one may
use the density operator $\rho$ of this preparation 
to compute the probability of an outcome $k$.  We have that
$\Pr(K = k | \rho) = {\rm Tr}(\Pi_k \rho)$, where $\Pi_k = M_k^{\dagger}M_k$.
This trace is linear on $\Pi_k$ and linear on $\rho$. Therefore,
it is often advantageous to work with $\Pi_k$ and $\rho$ rather
than with $M_k$ and $|\psi_{\alpha}\rangle$.  
The matrix representation 
of the operator $\rho$ in the basis $\{|\psi_{\alpha} \rangle\}$ is
defined by $(\rho)_{\alpha,\alpha'} = 
\langle \psi_{\alpha} |\rho | \psi_{\alpha'} \rangle$.

In accordance with the BB84 coding scheme, 
the states $|0\rangle_+$, $|0\rangle_{\times}$, $|1\rangle_+$
and $|1\rangle_{\times}$ corresponds to one photon polarized
at $0^{\circ}$, $45^{\circ}$, $90^{\circ}$ and $-45^{\circ}$
degrees respectively.  Note  that $+$ and $\times$ corresponds
to the bases $\{|0\rangle_+, |1\rangle_+\}$ and 
$\{|0\rangle_{\times}, |1\rangle_{\times} \}$ respectively.
For every $\theta \in \{+,\times\}^n$ and every $w \in \{0,1\}^n$,  
$|\psi_{w,\theta} \rangle$ denotes the product
state $|w_1\rangle_{\theta_1} \ldots |w_n \rangle_{\theta_n}$.
For any set of positions $E = \{\gamma_1,\ldots, \gamma_N\}$, let
$w[E]$ be the string given by $w[E]_i = w_{\gamma_i}$,
$1 \leq i \leq N$, and
let  $|\psi_{w,\theta}[E]\rangle$ be  the product state
$|w_{i_1}\rangle_{\theta_{i_1}} \ldots |w_{i_N}\rangle_{\theta_{i_N}}$
for the photons with position in $E$.

\section{The String-QOT protocol and its security}
The QOT protocol considered by Yao in \cite{yao95} 
is a variant of the QOT protocol which
has been first proposed by Cr\'epeau \cite{crepeau87,crepeau90} and
improved later in \cite{bbcs92,crepeau94}. 
We consider the natural generalization
of this single bit QOT protocol to a string QOT.
In this String-QOT protocol, 
$n$ is the number of photons sent in the protocol, 
$b$ is the string sent by Alice,
$m$ is the length of $b$, 
$r$ is the number of redundant bits needed for error correction, 
and $N = \lfloor .24 n \rfloor$ is the length of the
string shared between Alice and Bob {\em before} privacy amplification. 

\begin{enumerate}
\item[] \hspace{-\labelwidth} STRING-QOT($b$) \vspace{.1in}
\item \label{Code} 
      Alice picks a random uniformly chosen $(r+m) \times N$ boolean matrix $f$
      where the $r$ first rows define a matrix $g$ used for error correction
      and the $m$ following rows define a matrix $h$ used for 
      privacy amplification (see step \ref{privacy}).
\item Bob picks a random uniformly chosen $\hat{\theta} 
      = \hat{\theta}_1 \ldots \hat{\theta}_n \in \{+,\times\}^n$
      and makes a quantum commit of all $\hat{\theta}_i$ to Alice.
\item Alice picks a random uniformly chosen $w \in \{0,1\}^n$, 
      a random uniformly chosen $\theta \in \{+, \times\}^n$,
      and sends to Bob $n$ photons in the state $|\psi_{w,\theta}\rangle$.
\item \label{first_measurement}
      Bob measures every photon $i$ in basis $\hat{\theta}_i$, record the
      results $\hat{w}_i$ and makes a quantum commit of all $n$ bits
      $\hat{w}_i$ to Alice.
\item \label{test}
      Alice picks a random uniformly chosen subset 
      $R \subseteq \{1,\ldots, n\}$
      and tests the commitment made by Bob at positions $i \in R$. 
      If more than $\delta n$ positions $i \in R$ reveal 
      $\theta_i = \hat{\theta}_i$
      and $w_i \neq \hat{w}_i$, then Alice stops the protocol; otherwise,
      the test result is accepted. 
\item \label{announce} 
      Alice announces the string $\theta$.  
      Let $T_0$ be the set of all $i$ with $\theta_i = \hat{\theta}_i$, 
      and let $T_1$ be the set of all $i$ with 
      $\theta_i \neq \hat{\theta}_i$.  
      Bob chooses a set $E_0 \subseteq T_0 - R$, a set
      $E_1 \subseteq T_1 - R$, where $|E_0| = |E_1| = N$,  
      and  announces $\{E_0,E_1\}$ in random order to Alice.  
\item \label{privacy} 
      Alice chooses at random a set $E_c \in \{E_0,E_1\}$.
      For error correction, she announces the matrix $g$ and
      the string $s = g \, w[E_c]$. For the computation of $b$,
      she announces the matrix $h$ 
      and the string $a = b \oplus ( h\, w[E_c] )$.
\item \label{final}
      If $c = 0$,  Bob obtains $w[E_c]$ by 
      correcting the errors in $\hat{w}[E_c]$, then
      he computes the intermediary string $t = h \, w[E_c]$
      and obtains the string $b$ via $b = a \oplus t$. 
      If $c = 1$, Bob obtains no information about $t$ 
      and, thus, no information about $b$.  
\end{enumerate}
Yao's QOT protocol is exactly as above, except that 
$r = 0$, $m = 1$ and the $1 \times N$ matrix $f$ is 
$(1,1,\ldots,1)$, that is, there is no error correction and 
there is only one secret bit $t = t_1$ which is the exclusive
or of all the bits in $w[E_c]$.

The QKD version is identical to 
the String-QOT protocol, except that Bob announces 
$E_0$ to Alice rather than $\{E_0, E_1\}$ 
and Alice always chooses $c = 0$.
In this paper, we shall only consider attacks that correspond
to attacks that may be executed by Eve in the QKD version.  
Clearly, Eve has no control over the set $E_0$ (and $E_1$), so
we shall assume that Bob constructs $E_0$ and $E_1$
as specified in the protocol. The case in which
there is no restriction on $E_0$ and $E_1$
is not more difficult, but we don't need it
to obtain the security of the QKD protocol.

In most cases, a random variable is represented by an upper case
letter, whereas the value taken by such a variable is represented by
a lower case letter, for instance, the bit $c$ is the value taken
by  a random variable $C$.  However, if the value itself is represented
by an upper case letter which is typically the case when the
value is a set, we use bold face typesetting
for the random variable to distinguish it from its value.  

Let $V$ be Bob's view at the end of the protocol.  
Let ${\it Pass}$ be the binary random variable that takes the
value  $1$ if and only if the test result is accepted. 
To  obtain the security of the above 
protocol against Bob, for any attack where $E_0$ and $E_1$ are honestly
chosen, we show that there exists 
a factor of security $\xi > 0$ such that,
for any initial distribution of probability on $B$,
$I(B;V | {\it Pass} = 1 \wedge C = 1 )
\times \Pr({\it Pass} = 1) \leq 2^{-\xi  n}$.

\section{Bob's view}
Let us assume that
the possible values $(b, w,\theta)$ of $(B,W,\Theta)$ are stored
in orthonormal states $|b,w,\theta\rangle_{\cal C}$.  
The entire view of Bob can be seen as the outcome of a 
measurement executed 
on $|b, w,\theta\rangle_{\cal C} |\psi_{w,\theta}\rangle$.  This 
measurement is not executed by Bob alone.  For instance,
the announcement of $\theta$ by Alice is part of this
measurement.
Furthermore, we shall generously assume that  at the end
Alice announces $w[\bar{E}_c]$ to Bob.
 
Let us  analyze  the operation $M_v$ associated with
a view $v$.  We consider a fixed value of $\hat{\theta}$.
At step \ref{first_measurement} the measurement operates 
only on $|\psi_{w,\theta}\rangle$
and returns $\hat{w}$: we consider
the classical computation of $\hat{w}$
as part of the measurement executed by a dishonest Bob.
The corresponding operation on the photons is denoted 
$M_{\hat{w}}$. At step  \ref{test},  $R$ is
chosen  by Alice and announced to Bob.  This has
no physical effect on the initial state, but still the corresponding
operation is $M_R = 2^{-n}\, {\bf I}$. Next, Alice announces
the result of the test.  This corresponds to a projection
$P_{\it pass}$ on the classical part of the state space.
Note that this projection is defined in view of 
$\hat{w}$ which is obtained from a measurement on the photons.  
At step \ref{announce} Alice announces $\theta$.
The corresponding  operation 
is  the projection $P_{\theta} = 
|\theta\rangle \langle \theta |_{\cal C}$.
The announcement of $E_c$ corresponds to the
operation $M_c = 2^{-1}\, {\bf I}$.
Similarly, let $P_s$ and $P_a$ be respectively 
the projection that corresponds to the announcement
of $s$ and $a$.  We have that $P_s$ projects on the span of the states 
$|w[E_c]\rangle_{\cal C}$ such that $S = s$ and
$P_a$ projects on the 
span of the states $|b, w[E_c]\,\rangle_{\cal C}$ such that 
$A = T(w[E_c]) \oplus b = a$.
Note that, because Bob could have some
initial information about $b$, the condition $A = a$ may actually
provide information about $t = b \oplus a$.
Finally, let $P_w$ be the projection
$|\, w[\bar{E}_c] \, \rangle \langle\,  w[\bar{E}_c] \, |_{\cal C}$
which corresponds to the announcement of $w[\bar{E}_c]$.  

Note that Bob has no advantage in measuring the
photons at step \ref{announce} 
(because he creates $E_0$ and $E_1$ honestly). So the
operation $M_{\hat{w}}$ 
on the photons at step \ref{test} remains the same
at step \ref{announce}.  At step \ref{privacy},
Alice announces the information for privacy amplification and 
error correction, but this is under Alice's control
and operates only on the classical part of the
initial state.  
Certainly, at step \ref{final},  Bob is free to
execute on the residual state of the photons 
the complete measurement of his choice.  
The final operation on the initial state 
$|b, w,\theta\rangle_{\cal C} |\psi_{w,\theta}\rangle$ is of the form
$M_v = 2^{-(n+1)} P_{\cal C} |v\rangle\langle \phi_v |$
where $|v\rangle \langle \phi_v|$ operates on 
$|\psi_{w,\theta}\rangle$ 
and  $P_{\cal C}$ is the projection $P_w P_a P_s P_{\theta}$
on the classical part $|b, w,\theta\rangle_{\cal C}$.
The projection $P_{\it pass}$ does not appear because
it is implicit in $P_w P_{\theta}$. 
   
\section{The small distance property} \label{small_distance_section}
In this section, we want to find a property on $M_v$ that can be
proven  using  the fact that Bob must pass the test.
Of course, we also want a property that implies
that Bob has no information when $c = 1$.  
We recall that
no more than $\delta n$ positions $i$ for which
$\theta_i = \hat{\theta}_i$ and $w_i \neq \hat{w}_i$
are tolerated in the test.

Let us consider an example in which  Bob stores some photons and 
measures them only after that the bases have been announced
by Alice.  Let $\epsilon = 8 \delta$.  Bob cannot store 
much more than $\epsilon n$ photons, 
because otherwise he will not pass the test: half of the
photons are used for the test, 
half of these tested photons will be in the correct
basis and half of these will create an error.  Consider the
case where Bob stores exactly $\epsilon n$ photons.  
Let $F$ be the set of stored photons and
$\bar{F}$ the set of non stored photons.
To pass the test, Bob
measures the non stored photons using the committed 
string of bases $\hat{\theta}[\bar{F}]$ and
obtains $\hat{w}[\bar{F}]$.  After that
he has learned all the classical information  
that Alice announces, Bob measures the stored photons in 
the correct bases $\theta[F]$ and obtains $w[F]$. 
The value $(\hat{w}, \hat{\theta}, \theta)$ is
fixed in the final view $v$ and
the corresponding vector is 
$|\phi_v \rangle 
= |\psi_{\hat{w}, \hat{\theta}}[\bar{F}] \rangle  
| \psi_{w,\theta}[F]\rangle $.

In which way the dishonest vector $|\phi_v\rangle
= |\psi_{\hat{w}, \hat{\theta}}[\bar{F}] \rangle  
| \psi_{w,\theta}[F]\rangle$ 
is close from the honest vector
$|\phi_v \rangle = |\psi_{\hat{w},\hat{\theta}} \rangle$\,?
If we expand the state 
$|\psi_{\hat{w}, \hat{\theta}}[\bar{F}] \rangle  
| \psi_{w,\theta}[F]\rangle$ in the
basis $\{|\psi_{\hat{w}, \hat{\theta}} \rangle \}$,
we obtain 
$|\psi_{\hat{w}, \hat{\theta}}[\bar{F}] \rangle  
| \psi_{w,\theta}[F]\rangle 
= \sum_{\alpha} \lambda_{\alpha} 
|\psi_{\alpha, \hat{\theta}} \rangle$ 
where $\lambda_{\alpha} \neq 0$ only if 
we have $\alpha[\bar{F}] = \hat{w}[\bar{F}]$.
In particular, $\lambda_{\alpha} \neq 0$ implies 
$d(\alpha,\hat{w}) \leq \epsilon n$.
Of course, Bob could choose the photons that he stores at random
and in view of the previous outcomes.  In this case,
we cannot expect that, for some fixed set $F$,
$\lambda_{\alpha} \neq 0$
implies 
$\alpha[\bar{F}] = \hat{w}[\bar{F}]$. 
However, it is still reasonable
to expect that $\lambda_{\alpha} \neq 0$ implies 
$d(\alpha,\hat{w}) \leq \epsilon n$.  That is,
the state $|\phi_v\rangle$ must be in the span
of the states $|\psi_{\alpha,\hat{\theta}}\rangle$
with $d(\alpha,\hat{w}) \leq \epsilon n$.  
This is exactly the property that is called
the low weight property by Yao~\cite{yao95}.
In Yao's proof, $\epsilon = 1/40$.
The test of the QOT protocol
in Yao's proof tolerates no error at all: $\delta = 0$.
However, Yao's proof works exactly in the same way even
when $\delta > 0$.  In section \ref{proving_it} 
we shall briefly sketch an alternative proof.  

Let us formulate
the low-weight property in terms of $M_v$ and the set $E_c$.
We consider $E_c$ because it
contains the relevant positions.
Let $E \subseteq \{1,\ldots, n\}$ be any set of
positions and $\epsilon$ be some small positive number.
Let $d_E(\alpha, \alpha') = \#\{i \in E\;|\; \alpha_i \neq \alpha'_i \}$.
If $E = \{1,\ldots, n\}$, then $d_E(\alpha,\alpha')$ is the usual Hamming
distance. 
We denote $L_1[E,\epsilon n]$  the span of 
the states $|\psi_{\alpha,\hat{\theta}}\rangle$
where  $d_E(\alpha, \hat{w}) \leq \epsilon n$.  
We denote $L_0[E,\epsilon n]$ the span of the states 
$|\psi_{\alpha,\hat{\theta}}\rangle$
where $d_E(\alpha,z) > \epsilon n$. 
We denote $P_j[E,\epsilon n]$ the projection
on $L_j[E,\epsilon n]$.

Let $P_0 = P_0[E_c, \epsilon n]$ and $P_1 = P_1[E_c,\epsilon n]$.
A vector $|\phi\rangle$ in the state space of the photons
has the $\epsilon n$-small
distance property if and only if $P_0 |\phi\rangle = 0$.
In other words, it must be in $L_1[E_c, \epsilon n]$.
The operation 
$M_v$ has the $\epsilon n$-small-distance property
if and only if, for every $(b,w,\theta)$, 
$M_v P_0\; |b, w,\theta \rangle_{\cal C} |\psi_{w,\theta}\rangle = 0$.
The small-distance property corresponds to what Yao calls
the low-weight property in \cite{yao95}.  
Note that 
Yao defines the low weight property in terms of all the
positions, not only those in $E_c$.  This difference is not so important: 
it is clear that $L_1[\{1,\ldots,n\},\epsilon n]$
is a subspace of $L_1[E_c, \epsilon n]$,
so Yao's low-weight property implies
the small distance property.

\section{Using the small distance property}
We now show that if the small distance property
holds and $c = 1$, then $v$ provides no
information at all on $b$.  This corresponds to
a generalization of lemma $1$ in Yao's paper~\cite{yao95}.
The minimum  distance of a code $C$ is the minimum
Hamming distance $d(c,c')$ where $c$ and $c'$ are distinct codewords in $C$.
Let $C_0^{\perp}$ be the span of the $(r+m)$ rows of the matrix $f$ 
seen as vectors in ${\rm GF}(2)^N$. 
Let $d N$ be the minimum distance of $C_0^{\perp}$. 
Because the matrix $f$ is chosen at random, for any $\eta > 0$, 
except with negligible probability, 
we have $d > H^{-1}(1 - \frac{r+m}{N} ) - \eta$, 
where $H(x) = - (\;x \,\lg(x) + (1-x)\,\lg(1-x) \;)$.     
\begin{lemma} \label{main_lemma}
If  $\epsilon n < \frac{d N}{2}$, $c = 1$ 
and $M_v$ has the 
$\epsilon n$-small distance property, then the outcome $v$ 
provides no information at all on the string $b$.  
\end{lemma}
\begin{proof}
The basic idea is to show that, for a fixed $v$ such that $c = 1$,
the probability of $V = v$  given $B = b$, denoted $p(v|b)$, 
is the same for all $b$. 
For every $(w',\theta')$, let $p(v|b, w',\theta') 
= \Pr(V= v | B = b \wedge W = w \wedge \Theta = \theta')$.
We have that 
$p(v|b) = 4^{-n} \sum_{w',\theta'} p(v|b,w',\theta')$.
Now, let ${\cal P}_{v,b}$ be the set of pair $(w' ,\theta')$
such that  
\begin{equation} \label{classical_constraint}
P_{\cal C} |b,w',\theta'\rangle_{\cal C} \neq 0.
\end{equation}
Equation (\ref{classical_constraint})
must hold if we want to have $p(v|b,w',\theta') \neq 0$.  
Since, we are only interested in $(w',\theta')$
that contributes to $p(v|b)$, 
in what follows we only consider the pair $(w',\theta')$ 
in ${\cal P}_{v,b}$.  
We obtain that $P_{\cal C}$ operates as the identity operator 
on  $|b,w',\theta'\rangle_{\cal C}$. 
Furthermore, one may easily check that
(\ref{classical_constraint})
implies that we can express the $\epsilon n$-small distance 
property on $M_v$ via the following
equation.  
\begin{equation} \label{small_distance}
\langle \phi_v | P_0 | \psi_{w',\theta'}\rangle = 0.
\end{equation}
Because of these two facts, from hereafter we can
ignore the classical  part  of the initial state 
in our  computation.  
Now, equation (\ref{classical_constraint}) 
implies $w'[\bar{E}_c] = w[\bar{E}_c]$, 
$\theta' = \theta$, $g\, w[E_c] = s$ and $h\, w[E_c] = t = b \oplus a$.
The two last constraints can be written in one equation
$f\, w[E_c] = x$ where $x$ is the concatenation of $s$ and $t$.    
The only degree of freedom is $\beta \stackrel{def}{=} w'[E_c]$  
restricted by $f \beta = x$.
Let $C_x = \{ \beta \in \{0,1\}^N \; | \; f \beta = x\}$.
There is a one-to-one correspondence between
the strings $\beta \in C_x$ and the pairs 
$(w',\theta') \in {\cal P}_{v,b}$.
Let $p(v|\beta) = p(v | b, w', \theta')$
and  $|\psi_{\beta,\theta}\rangle = |\psi_{w',\theta'}\rangle$.
Ignoring the classical part
of the initial state and  using  (\ref{small_distance}) 
we obtain  $p(v | \beta ) = 
|\langle \phi_v | \psi_{\beta, \theta} \rangle|^2 
 =  |\langle \phi_v | P_0 + P_1 | \psi_{\beta, \theta}\rangle|^2
 =  |\langle \phi_v | P_1 | \psi_{\beta, \theta}\rangle|^2.$

Now, we would like to restrict our analysis to
the photons with position in $E_c$.  
One may insert the projection 
$P = | \, \psi_{w,\theta}[E_c]\,\rangle \langle\, \psi_{w,\theta}[E_c]\,|$ 
in front  of  the state $| \psi_{\beta,\theta}\rangle $ because
this projection is implicit in the definition of this state.
One obtains 
$p(v|\beta) = |\langle \phi_v | P_1 P |\psi_{\beta,\theta}\rangle|^2$.
These two projections commute, so we obtain
$p(v|\beta) = |\langle \phi'_v | P_1 |\psi_{\beta,\theta} \rangle|^2$
where $|\phi'_v \rangle = P |\phi_v\rangle$.
Note that 
$|\phi'_v\rangle = |\psi_{w,\theta}[\bar{E}_c]\rangle|\phi''_v\rangle$
and $|\psi_{\beta, \theta} \rangle =  
|\psi_{w,\theta}[\bar{E}_c]\rangle|\tilde{\psi}_{\beta,\theta} \rangle$
where both $|\phi''_v\rangle$ and $|\tilde{\psi}_{\beta,\theta} \rangle$
are states for the photons with position in $E_c$. We obtain that
$p(v | \beta) =
|\langle \phi''_v | P_1 | \tilde{\psi}_{\beta,\theta}  \rangle|^2
= |\langle \tilde{\phi}_v | \tilde{\psi}_{\beta,\theta} \rangle|^2$
where $|\tilde{\phi}_v\rangle =  P_1 |\phi''_v\rangle$ has 
the $\epsilon n$-small-distance property.
Now, consider the density operators
$
\rho_x = 2^{-k} \sum_{\beta \in C_x} \;
|\tilde{\psi}_{\beta,\theta} \rangle\langle \tilde{\psi}_{\beta,\theta} |
$ 
where $k = N - r - m$.
We shall show that these density operators cannot be
distinguished by any state $|\tilde{\phi}\rangle$ that has
the $\epsilon n$-small distance property.
In section \ref{density_matrices}, 
it is shown that, in the context $E_c = E_1$, for every
$\beta \in C_x$, the matrix
representation of $\rho_x$ in Bob's basis 
$\{ | \tilde{\psi}_{\alpha,\hat{\theta}} \rangle
\; | \; \alpha \in \{0,1\}^N \} $
is given by
\[
(\rho_x)_{\alpha,\alpha'} = 2^{-N} \times 
\left\{\begin{array}{ll}
0 & \mbox{ if } (\alpha \oplus \alpha') \not \in C_0^{\perp} \\
(-1)^{(\alpha \oplus \alpha') \otimes \beta} & \mbox{ otherwise }
\end{array} \right.
\]
For every pair  of distinct strings $x,x' \in\{0,1\}^{m+r}$, 
we have that a necessary condition for
$(\Delta \rho)_{\alpha,\alpha'} =
(\rho_x)_{\alpha,\alpha'} - (\rho_{x'})_{\alpha,\alpha'} \neq 0$
is that $(\alpha \oplus \alpha')$ belongs to  $C_0^{\perp}$
and is different from $0$.   Therefore, a necessary condition 
for $(\Delta \rho)_{\alpha,\alpha'} \neq 0$  is
that $d(\alpha, \alpha') > dN$.  Therefore, for
every $(\alpha, \alpha')$ such that
$(\Delta \rho)_{\alpha,\alpha'} \neq 0$, 
one of 
$|\psi_{\alpha,\hat{\theta}} \rangle$ or
$|\psi_{\alpha',\hat{\theta}} \rangle$
 belongs
to $L_0[E,\epsilon n]$.  We obtain
\[
\langle \phi | \Delta\rho | \phi \rangle
 = \sum_{\alpha,\alpha'} (\Delta \rho)_{\alpha,\alpha'}  
\langle \phi |\tilde{\psi}_{\alpha,\hat{\theta}}\, \rangle 
\langle\tilde{\psi}_{\alpha',\hat{\theta} } \, | \phi \rangle = 0
\] 
This concludes the proof. \qed
\end{proof}

\section{The density matrices}  \label{density_matrices}
In this section, we consider only the photons
with positions in $E_1 = E_c$.   
Therefore $\hat{\theta}$ is the {\em opposite} of $\theta$, 
that is, $(\forall i)\, \hat{\theta}_i \neq \theta_i$.  
We temporarily remove the tilde over the symbol $\psi$.  
It is as if we considered
the general situation where $N$ photons are
sent from Alice to Bob in a string of bases 
$\theta \in \{+,\times\}^N$  and 
we want to find the matrix representation of the density operators
$
\rho_x = 2^{-k} \sum_{\beta \in C_x } 
|\psi_{\beta, \theta } \rangle \langle \psi_{\beta, \theta } |
$
in the opposite basis $\{ |\psi_{\alpha,\hat{\theta} }\rangle \}$.
We need some basic tool.  For every vector $\beta \in {\rm GF}(2)^N$, 
the mapping $\beta' \mapsto \beta' \oplus \beta$
on ${\rm GF}(2)^n$ corresponds  
to a unitary transformation $U_{\beta}$ 
on the state space of the photons
defined via $U_{\beta} |\psi_{\beta',\theta} \rangle = 
|\psi_{\beta \oplus \beta' ,\theta}\rangle$. 
One may easily check that, for every 
position $i$ where $\beta_i = 1$, 
the transformation $U_{\beta}$ maps $|0\rangle_{\hat{\theta}_i}$ into itself 
and $|1\rangle_{\hat{\theta}_i}$ into $- |1\rangle_{\hat{\theta}_i}$.  
So,  if there is an even number of 
positions $i$ where $\alpha_i = \beta_i = 1$, we have
$U_{\beta} |\psi_{\alpha,\hat{\theta} }\rangle 
= |\psi_{\alpha,\hat{\theta}}\rangle$,  otherwise,
we have $U_{\beta} |\psi_{\alpha,\hat{\theta}}\rangle
= - |\psi_{\alpha,\hat{\theta}}\rangle$.
In terms of the operation $\odot$ on 
the vector space ${\rm GF}(2)^n$, we have 
\[
U_{\beta} |\phi_{\alpha,\hat{\theta}}\rangle =  \left\{ 
\begin{array}{ll}  
|\psi_{\alpha,\hat{\theta} }\rangle & \mbox{if } \beta 
  \odot \alpha = 0 \\
-|\psi_{\alpha,\hat{\theta} }\rangle &  \mbox{if } \beta 
  \odot \alpha = 1 
\end{array} \right.
\]
For every $\beta \in C_x$, 
we have $C_x = C_{\bf 0} \oplus \beta$.
Therefore, for every $\beta \in C_x$, 
\begin{equation} \label{rho_x}
\rho_x =  U_{\beta} \rho_{\bf 0} U_{\beta},
\end{equation} where we have used 
$U_{\beta}^{\dagger} = U_{\beta}$.  
For any operator $\rho$ and any $\beta$, 
one may easily check that, in Bob's basis, 
\begin{equation}\label{translation}
(U_{\beta} \rho U_{\beta})_{\alpha, \alpha'}
= (-1)^{(\alpha \oplus \alpha') \odot \beta} \times (\rho)_{\alpha,\alpha'}.
\end{equation}
Therefore, in view of (\ref{rho_x}) and (\ref{translation}),
we are done if we have the  matrix representation of 
the density operator $\rho_{\bf 0}$ in Bob's basis.  

Let $k = N-m-r$ and $\{\beta_1, \ldots, \beta_k\}$  be a basis of $C_0$.
For every $j = 1, \ldots, k$, let
$C^{(j)}$ be the span of $\{\beta_1, \ldots, \beta_j\}$
and $\rho^{(j)}
= 2^{-j} \sum_{\beta \in C^{(j)} } | \psi_{\beta,\theta} \rangle
\langle \psi_{\beta,\theta} |.$
Note that $\rho_{\bf 0} =  \rho^{(k)}$ and $C_{\bf 0} = C^{(k)}$. 
We shall show by induction on $j$, that for 
$j = 0, \ldots, k$,
\begin{equation} \label{rho_j}
(\rho^{(j)} )_{\alpha,\alpha'} = 2^{-N} \times
\left\{ \begin{array}{ll}
0 & \mbox{ if } (\alpha \oplus \alpha') \not \in C^{(j)\perp} \\
1 & \mbox{ otherwise}
\end{array} \right.
\end{equation}
The case $j= 0$ can be easily computed: $C^{(0)} = \{0\}$
and $C^{(0)\perp} = {\rm GF}(2)^n$.  We assume
that (\ref{rho_j})  holds for $j$ and obtain it for $j+1$.
Because $C^{(j+1)} = C^{(j)} \cup ( C^{(j)} \oplus \beta_{j+1} )$,
we have that
\begin{equation} 
\rho^{(j+1)}_{\bf 0} = 1/2 ( \rho^{(j)}_{\bf 0} + 
              U_{\beta_{j+1}} \rho^{(j)}_{\bf 0} U_{\beta_{j+1} } ).
\end{equation}
Therefore, using formula \ref{translation}, we obtain
\[
( \rho^{(j+1)} )_{\alpha, \alpha'} =
1/2 (\rho^{(j)})_{\alpha,\alpha'}
(1 - (-1)^{ (\alpha \oplus \alpha') \odot \beta_{j+1} } ).
\]
Note that $(\rho^{(j+1)})_{\alpha, \alpha'}$ is either $0$ or $2^{-N}$.  
We obtain that 
$(\rho^{(j+1)})_{\alpha, \alpha'} = 2^{-N}$ 
if and only if $(\rho^{(j)})_{\alpha,\alpha'} \neq 0$
and $(\alpha \oplus \alpha') \odot \beta_{j+1} = 0$.
So,   $(\rho^{(j+1)})_{\alpha, \alpha'} = 2^{-N}$ 
if and only if, for every $\beta \in C^{(j+1)}$,
$(\alpha \oplus \alpha') \odot \beta = 0$.  This
last condition is 
equivalent to $(\alpha \oplus \alpha') \in C^{(j+1)\perp}$.
This concludes the induction.
Using the density matrix of $\rho_{\bf 0} = \rho^{(k)}$, together with 
formula \ref{rho_x} and \ref{translation}, we finally 
obtain that, for every $\beta \in C_x$,
\[
(\rho_x)_{\alpha,\alpha'} = 2^{-N} \times 
\left\{\begin{array}{ll}
0 & \mbox{ if } (\alpha \oplus \alpha') \not \in C_0^{\perp} \\
(-1)^{(\alpha \oplus \alpha') \otimes \beta} & \mbox{ otherwise }
\end{array} \right.
\]

\section{Proving the small distance property} \label{proving_it}
Consider an example where Bob chooses a 
random bit ${\it OK}$ and stores all the photons 
when and  only when ${\it OK} = 1$. In this case,
Bob passes the test with a
probability a little bit greater than $1/2$
and the small distance property
holds with probability $1/2$.  The point is that we should
not expect that, if Bob has a significant
probability to pass the test, then the small
distance property always holds.   In this example,
except with
negligible probability, the small distance property 
holds when Bob passes the test.

Consider another example where Bob commits
$\hat{\theta} = +^n$,
measures every photon in a fixed basis $\theta'$
and commits the outcome $\hat{w}$. The 
fixed basis $\theta'$ cannot be too far away from $+$
because otherwise Bob will not pass the test. 
Without loss of generality, 
assume that the magnitude of
$\mbox{}_+\!\langle 0|0\rangle_{\theta'}
= \mbox{}_+\!\langle 1|1\rangle_{\theta'} = c_{\theta'}$
is close to $1$ and the magnitude of
$\mbox{}_+\!\langle 0|1\rangle_{\theta'} =
\mbox{}_+\!\langle 1|0\rangle_{\theta'} = s_{\theta'}$ is close to $0$.
The value $\hat{w}$ is included in $v$ and 
$|\phi_v\rangle = |\psi_{\hat{w},\theta'}\rangle$.
If we expand $|\phi_v\rangle$  in Bob's basis $+^n$ we obtain
$|\phi_v \rangle = \sum_{\alpha}
\langle \psi_{\alpha, +^n} | \psi_{\hat{w},\theta'} \rangle 
|\psi_{\alpha, +^n} \rangle$.
Note that 
$|\langle \psi_{\alpha, +^n} | \psi_{\hat{w},\theta'} \rangle |
= |s_{\theta'}|^{d(\alpha,\hat{w})} \times
  |d_{\theta'}|^{n - d(\alpha,\hat{w})}$.
So $| \lambda_{\alpha}|  =
| \langle \psi_{\alpha, +^n} | \psi_{\hat{w},\theta'} \rangle | $
is very small when $d(\alpha,\hat{w})$ is large.
In this second example, the small distance property
does not hold, but it {\em almost} holds.

Now, we briefly sketch a proof that,
for every strategy used by Bob,
except with negligible probability, 
if Bob passes the test, then the small distance property {\em almost}
holds. A complete proof is found in \cite{yao95}.  
Let $\gamma = 10^{-6}$ and ${\it Info}$ be the binary random variable
that takes the value $0$ if and only if 
\[
\| M_v  P_0  |\psi_{w,\theta} \rangle \|^2
\leq 2^{-\gamma n} \| M_v | \psi_{w,\theta} \rangle \|^2.
\]
The condition ${\it Info} =0$ means 
that, for all practical purposes, we can use the
small distance property, obtain (\ref{small_distance}), 
etc.\  in our proof of lemma \ref{main_lemma}.  

So, we want to obtain that 
if $\Pr({\it Pass} = 1) > 2^{-\gamma n}$  then 
\begin{equation} \label{End}
\Pr({\it Info} = 1 \; |\; {\it Pass} = 1 ) \leq 2^{-\gamma n}.
\end{equation}
The variable ${\it Info}$ concerns the final view of Bob.
It is easier to
consider the situation just after the announcement of $\theta$.
Therefore, let us consider the ratio
\[ 
r({\it pass}, \theta,R,\hat{w}) 
= \frac{{\rm Tr}(P_0 \, \Pi_{({\it pass}, \theta,R, \hat{w})} \, P_0\, \rho)}
             {{\rm Tr}(\Pi_{({\it pass}, \theta, R,\hat{w})} \, \rho)}      
\]
where $\rho$ is Alice's preparation and
$\Pi_{({\it pass},\theta,R, \hat{w})} = 
 M_{(\theta,{\it pass}, R,\hat{w})}^{\dagger} \, 
M_{(\theta,{\it pass}, R,\hat{w})}$.
We shall briefly sketch why
$\Pr({\it Pass} = 1) > 2^{-2\gamma n}$
implies that
\begin{equation} \label{After_test} 
\langle r({\it pass}, \theta,R,\hat{w} ) \rangle_{{\it Pass}\, = 1} 
\leq 2^{-2\gamma n}
\end{equation}
where $\langle r \rangle_{{\it Pass}\, = 1}$ denotes the
expected value of $r$ in the context ${\it Pass} = 1$. 
This do the job because $\Pr({\it Pass} = 1 ) > 2^{-\gamma n}$
implies that  $\Pr({\it Pass} = 1 ) > 2^{- 2 \gamma n}$ and
expanding the expected value
$\langle r({\it pass}, \theta,R,\hat{w}) \rangle_{{\it Pass}\, = 1} $
and after some algebra, one obtains
that (\ref{After_test}) implies  (\ref{End}).  
One may check that
\begin{eqnarray}
{\rm Tr}(\Pi_{({\it pass},\theta, R,\hat{w})} \, \rho) 
   & = & p({\it pass}, \theta, R, \hat{w})  \nonumber \\
   &= & 8^{-n} \sum_{\alpha} \|\, M_{\hat{w} }\,
   P_{\it pass}[T_0 \cap R, \delta n]\, 
   |\psi_{\alpha, \hat{\theta}} \rangle\, \|^2  \label{denominator} \\
{\rm Tr}(P_0 \, \Pi_{({\it pass},R, \theta,\hat{w})} \, P_0\, \rho)
& = & 8^{-n} \sum_{\alpha} \| \,  M_{ \hat{w}} \,
P_{\it pass}[T_0 \cap R, \delta n] \, P_0 \,
|\psi_{\alpha, \hat{\theta} } \rangle \, \|^2  \label{numerator}
\end{eqnarray}
where $P_{\it pass}[T_0 \cap R, \delta n]$ 
refers to  section \ref{small_distance_section}.
The right hand side of (\ref{denominator}) and
(\ref{numerator}) can also be
obtained from the following definition of  $ {\it Pass}, \Theta, {\bf R}$
and $\hat{W}$.
Alice chooses $\theta$ and $R$ as usual, but prepares a perfectly
random state $|\psi_{\alpha, \hat{\theta}} \rangle$
using $\hat{\theta}$ rather than $\theta$.
Bob measures in the bases $\hat{\theta}$  
to obtain $\alpha$ and then executes $M_{\hat{w}}$ to 
obtain $\hat{w}$.  Finally, Alice announces $R$ and $\theta$.
Let $J[E, \tau n] = 0$ if and only if 
$d_E(\alpha, \hat{w}) \leq \tau n$,
and let ${\it Pass} = J[T_0 \cap R, \delta n]$.
The values of (\ref{denominator}) and (\ref{numerator})
are respectively  
$\Pr( \Theta = \theta \, \wedge \, 
J[T_0 \cap R, \delta n] = {\it pass} \, \wedge \,
{\bf R} = R, \wedge \, \hat{W} = \hat{w})$
and 
$\Pr(J[E_c,\epsilon] = 0 \, \wedge \, \Theta = \theta \, \wedge \, 
J[T_0 \cap R, \delta n] = {\it pass} \, \wedge \,
{\bf R} = R, \wedge \, \hat{W} = \hat{w})$.
Equation \ref{After_test} simply means
that $\Pr(J[E_c,\epsilon] = 0 \; | \; J[T_0 \cap R,\delta] = 1)
\leq 2^{-2\gamma n}$.  So, it is sufficient to show
$\Pr(J[E_c,\epsilon] = 0 \wedge J[T_0 \cap R,\delta] = 1)
\leq 2^{-4\gamma n}$.  For an appropriate $\epsilon > \delta$,
this is not hard to show. 
This concludes our sketchy proof of this section.

We are grateful to Eli Biham, Gilles Brassard, 
Claude Cr\'epeau, Christopher Fuchs, Tal Mor and Andrew Yao for 
fruitful discussions.  We especially thank
Tal Mor and Eli Biham for showing us preliminary version of
\cite{bms96} and a preliminary and partial version of \cite{bm96}.
These did not yet consider the density matrices approach for
the case  $r > 0$ or $m > 1$, but contained
the density matrices  for the case $r = 0$
and $m = 1$.  At the time, we also had these density matrices,
but the way they presented it helped us to make a guess on
the shape of the density matrices when $r>0$ and $m > 1$,
and this guess has been a great help in our computation. 
Our guess has also been proven independently in
later versions of \cite{bm96} in the context of the 
collective attack.

\end{document}